 \documentclass[final,3p,times,sort&compress]{elsarticle}

\usepackage{amssymb}
\usepackage{amsmath}
\usepackage{hyperref}
\usepackage{graphicx}
\usepackage{subcaption}
\usepackage{soul,color,xcolor}

\journal{Nuclear Physics B}

\begin{document}

\begin{frontmatter}

\title{Thermodynamics of high order correction for Schwarzschild-AdS black hole in non-commutative geometry}

\author{Baoyu Tan\corref{cor1}}
\ead{2022201126@buct.edu.cn}

\address{College of Mathematics and Physics, Beijing University of Chemical Technology,
	15 Beisanhuandonglu Street, Beijing, 100029, China}
\cortext[cor1]{Corresponding author}

\begin{abstract}
Under the premise that quantum gravity becomes non-negligible, higher-order corrections of non-commutative geometry dominate. In this paper, we studied the thermodynamics of high-order corrections for Schwarzschild-AdS black hole with Lorentz distribution in the framework of non-commutative geometry. Our results indicate that when high-order corrections dominate, the thermodynamic behavior of Schwarzschild-AdS black hole in non-commutative geometry will gradually approach that of ordinary Schwarzschild-AdS black hole. In addition, we also studied the Joule-Thomson effect of Schwarzschild-AdS black hole under high-order corrections.
\end{abstract}

\begin{keyword}
Black hole thermodynamics \sep Non-commutative geometry \sep Quantum gravity \sep Joule-Thomson effect
\end{keyword}

\end{frontmatter}

\section{Introduction}
Due to the fact that black holes are ideal physical objects for observing the contradiction between quantum theory and general relativity, researchers have conducted extensive research on quantum black holes and their dynamics in recent years \cite{fursaev1995temperature, xiao2022logarithmic, battista2024quantum, calmet2021quantum, wang2025dynamical, Munch_2023, Mele_2022, Bodendorfer_2019, Koch_2016, PhysRevD.97.024027, rincon2018scale, panotopoulos2021quasinormal}. Non commutative geometry provides new mathematical tools and physical perspectives for the study of quantum black holes, helping to understand issues such as the microstructure, entropy, and Hawking radiation of black holes.

Non-commutative geometry is an attempt to quantize spacetime \cite{panotopoulos2020quasinormal, frob2023noncommutative, smailagic2003feynman, smailagic2003uv, seiberg1999string, lizzi1999noncommutative, doplicher1995quantum, witten1986non}, which can be represented by the following algebra (we set $\hbar=1$):
\begin{equation}
	[x^\mu,x^\nu]=i\theta^{\mu\nu}.
\end{equation}
Where $\theta^{\mu\nu}$ is an anti-symmetric real tensor. The influence of non-commutative geometry on gravitational effects has been widely studied \cite{mann2011cosmological, modesto2010charged, nicolini2006noncommutative, lopez2006towards, calmet2006second, aschieri2005gravity, calmet2005noncommutative, chamseddine2001deforming}. Under the correction of non-commutative geometry, the mass distribution of Schwarzschild spacetime changes from point distribution to Gaussian distribution and Lorentz distribution \cite{araujo2024effects}:
\begin{equation}
	\rho^G(r)=\frac{M}{(4\pi\theta)^{3/2}}\exp(-\frac{r^2}{4\theta}).
\end{equation}
\begin{equation}
	\rho^L(r)=\frac{M\sqrt{\theta}}{\pi^{3/2}(r^2+\pi\theta)^2}.\label{eq:3}
\end{equation}
Researchers have found that the evaporation endpoint of Schwarzschild black hole with Gaussian distribution of mass is zero temperature extreme black hole \cite{nozari2008hawking}. The Schwarzschild black hole with a Lorentz distribution of mass has also been widely studied \cite{wang2024thermodynamics, campos2022quasinormal}.

In 1983, Hawking and Page proposed the phase transition of Schwarzschild-AdS black hole \cite{hawking1983thermodynamics}. Since then, thermodynamics of AdS spacetime has been extensively studied. In \cite{kastor2009enthalpy}, researchers corresponded mass of black holes to enthalpy and cosmological constant to pressure, deriving first law of thermodynamics for black holes in AdS spacetime. The most extensively studied AdS black hole is RN-AdS black hole. \cite{kubizvnak2012p} conducted in-depth research on the thermodynamic properties of RN-AdS black hole. Our previous work also investigated the thermodynamics of EGUP modified RN-AdS black hole surrounded by quintessence \cite{tan2025egup}. AdS black hole in modified gravity models, especially in Einstein-Gauss-Bonnet gravity, have been profoundly investigated \cite{PhysRevD.88.084045, cai2013pv}. Overall, the thermodynamic properties of AdS spacetime are very similar to those of Van der Waals systems \cite{kubizvnak2012p, rodrigues2022bardeen, li2022high, singh2020thermodynamics, magos2020thermodynamics, dolan2010cosmological, chamblin1999charged, chamblin1999holography}. In addition, if stress-energy tensor clearly contains mass of the black hole, the first law of black hole thermodynamics will be modified \cite{rodrigues2022bardeen, wang2024thermodynamics1, singh2020thermodynamics}.

In this paper, we calculated the high-order correction of Schwarzschild-AdS black hole thermodynamics with mass distribution of Lorentz distribution in non-commutative geometric framework. First, for small mass macroscopic black holes, since the radius of their event horizon is extremely small, the high order corrections in the non-commutative space clearly dominate, which has important practical significance. This is because the non-commutative space, by introducing non-commutative relations among coordinates, significantly changes the microscopic structure of space-time near the Planck scale. This change will lead to the non-commutativity of space-time, blur the point-structure in the classical sense, enhance quantum fluctuations, and make the high-order curvature terms non-negligible in the action. In addition, in the high curvature region near the event horizon, the non-commutative effect will amplify the quantum gravity corrections, making the high order terms that were originally secondary in the classical theory become dominant. More importantly, the unique ultraviolet-infrared correlation mechanism in the non-commutative field theory will cause the non-commutative parameters to transfer the high order corrections at high energy scales to the black hole thermodynamics at low energy scales. Even if the horizon radius of a macroscopic black hole is much larger than the Planck length, this mixed effect may still make the high order corrections significant.

This paper is organized as follows. In section \ref{sec:1}, we present high-order corrections to Schwarzschild-AdS black hole metric in non-commutative geometry and demonstrate that these corrections result in a decrease in the minimum radius of Schwarzschild-AdS black hole. In section \ref{sec:2}, we conducted a detailed study on the thermodynamics of high-order modified Schwarzschild-AdS black hole, providing the modified first law of thermodynamics, equation of state, phase transition behavior and criticality. We also calculated the black hole's heat capacity and Gibbs free energy. In section \ref{sec:3}, we calculated the Joule-Thomson effect under high-order correction and provided the inversion temperature curve. Finally, in section \ref{sec:4}, we summarized our results and discussed them. In this paper, we adopt the natural unit system $\hbar=k_B=G=c=1$.
\section{High order correction of Schwarzschild-AdS BH in non-commutative geometry}
\label{sec:1}
The Einstein field equation with cosmological constant is:
\begin{equation}
	R_{\mu\nu}-\frac{1}{2}g_{\mu\nu}R+\Lambda g_{\mu\nu}=8\pi T_{\mu\nu}.\label{eq:1}
\end{equation}
Where $R_{\mu\nu}$ is Ricci tensor, $g_{\mu\nu}$ is metric tensor, $R$ is Ricci scalar, and $T_{\mu\nu}$ is stress-energy tensor. The spherical metric can be represented by the following equation:
\begin{equation}
	\mathrm{d}s^2=-F(r)\mathrm{d}t^2+F^{-1}(r)\mathrm{d}r^2+r^2\mathrm{d}\Omega^2.\label{eq:2}
\end{equation}
Substituting (\ref{eq:2}) into (\ref{eq:1}) yields:
\begin{equation}
	F(r)=1+\frac{1}{r}\int_0^r(8\pi r^2T^0_0-r^2\Lambda)\mathrm{d}r.
\end{equation}
$T^0_0$ represents energy density, and according to the formula of Lorentz distribution in (\ref{eq:3}), we have:
\begin{equation}
	T^0_0=-\rho=-\frac{M\sqrt{\theta}}{\pi^{3/2}(r^2+\pi\theta)^2}.
\end{equation}
Where $M$ is mass of black hole and $\theta$ is non-commutative parameter. The line element of Schwarzschild-AdS black hole in non-commutative geometry is:
\begin{equation}
	F(r)=1-\frac{C}{r}+\frac{8M\sqrt{\theta}}{\sqrt{\pi}r^2}-\frac{16\sqrt{\pi}M\theta^{3/2}}{3r^4}-\frac{\Lambda}{3}r^2+\mathcal{O}(\theta^{5/2}).
\end{equation}
Where $C$ is integral constant. When $\theta\to 0$, Space time geometry returns to general Schwarzschild geometry, $C$ should be $2M$. Obviously, the correction of non-commutative geometry is a high-order correction, that is, term $\theta^{3/2}$.

To simplify the calculation, we introduce two parameters $\alpha$ and $\beta$:
\begin{equation}
	\alpha=\frac{8\sqrt{\theta}}{\sqrt{\pi}},~~~~\beta=\frac{16\sqrt{\pi}\theta^{3/2}}{3}.
\end{equation}
The line element of the Schwarzschild-AdS black hole with high-order correction in non-commutative geometry is:
\begin{equation}
	F(r)=1-\frac{2M}{r}+\frac{\alpha M}{r^2}-\frac{\beta M}{r^4}-\frac{\Lambda}{3}r^2.
\end{equation}
To ensure the existence of black holes, it is necessary to satisfy $F(r_H)=0$, $r_H>0$, $M>0$, $\Lambda<0$. So there are the following inequalities that depend on $\alpha$ and $\beta$:
\begin{equation}
	-2r_H^3+\alpha r_H^2-\beta<0.
\end{equation}
\begin{equation}
	r_H^4-2Mr_H^2-\alpha Mr_H^2-\beta<0.
\end{equation}
\section{Thermodynamics}
\label{sec:2}
Firstly, let's calculate some basic thermodynamic quantities. The mass of AdS black holes, also known as enthalpy, is determined by $F(r_H)=0$:
\begin{equation}
	M=\frac{\Lambda r_H^6-3r_H^4}{3\alpha r_H^2-6r_H^3-3\beta}.
\end{equation}
The entropy of black hole is Bekenstein-Hawking entropy, which is proportional to the area of event horizon:
\begin{equation}
	S=\frac{A}{4}=\pi r_H^2.
\end{equation}
For AdS black holes, pressure is proportional to the cosmological constant:
\begin{equation}
	P=-\frac{\Lambda}{8\pi}.\label{eq:4}
\end{equation}
The temperature of black hole is Hawking temperature, which is proportional to the surface gravity of event horizon:
\begin{equation}
	T=\frac{F^\prime(r_H)}{4\pi}=\frac{3\Lambda r_H^5-2\alpha\Lambda r_H^4-3r_H^3+3(\alpha+\beta\Lambda)r_H^2-6\beta}{6\pi\alpha r_H^3-12\pi r_H^4-6\pi\beta r_H}.\label{eq:5}
\end{equation}

\begin{figure}[h]
	\centering
	\includegraphics[width=10cm]{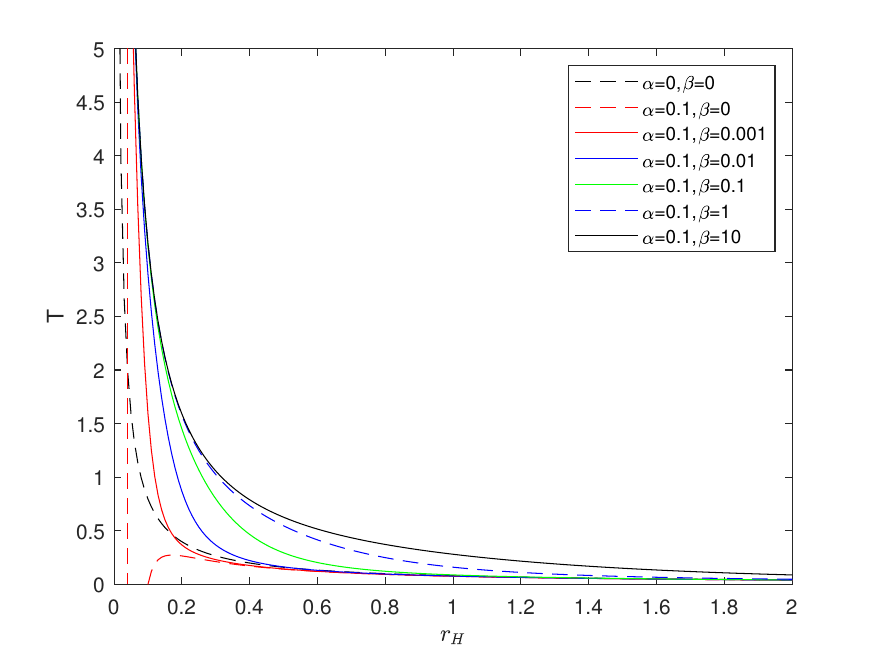}
	\caption{\label{fig:2}The Hawking temperature at $\alpha=0$, $\beta=0$ (black dashed line), $\alpha=0.1$, $\beta=0$ (red dashed line), $\alpha=0.1$, $\beta=0.001$ (red solid line), $\alpha=0.1$, $\beta=0.01$ (blue solid line), $\alpha=0.1$, $\beta=0.1$ (green solid line), $\alpha=0.1$, $\beta=1$ (blue dashed line), $\alpha=0.1$, $\beta=10$ (black solid line). We set $\Lambda=-0.01$.}
\end{figure}
In order to visually compare the cases without correction, only with leading order correction and with high-order correction, we have plotted the relationship between Hawking temperature and event horizon radius for these three cases in Figure \ref{fig:2} (we set $\Lambda=0.01$). The situation without correction corresponds to $\alpha=0$, $\beta=0$, only the leading order correction corresponds to $\alpha=0.1$, $\beta=0$, and the high-order correction corresponds to $\beta=0.001$, $\beta=0.01$, $\beta=0.1$, $\beta=1$, $\beta=10$ ($\alpha=0.1$), representing the change of high-order correction from secondary effect to primary effect. From Figure \ref{fig:2}, we can see that qualitatively speaking, the situation with high-order correction is consistent with the situation without correction.

When the stress-energy tensor clearly contains mass of black hole, first law of thermodynamics $\mathrm{d}M=T\mathrm{d}S+V\mathrm{d}P$ will be modified, which is evident:
\begin{equation}
	T\neq\left(\frac{\partial M}{\partial S}\right)_P,~~~~V\neq\left(\frac{\partial M}{\partial P}\right)_S.
\end{equation}
The modified first law of thermodynamics can be written as:
\begin{equation}
	\mathrm{d}\mathcal{M}=\mathcal{W}\mathrm{d}M=T\mathrm{d}S+V\mathrm{d}P.
\end{equation}
The correction factor $\mathcal{W}$ can be written as:
\begin{equation}
	\mathcal{W}=1+\int_{r_H}^{+\infty}4\pi r^2\frac{\partial T^0_0}{\partial M}\mathrm{d}r=1-\frac{a}{2r_H}+\frac{b}{2r_H^3}+\mathcal{O}(a^5).
\end{equation}
Using the modified first law of thermodynamics, we can obtain Hawking temperature and thermodynamic volume:
\begin{equation}
	T=\mathcal{W}\left(\frac{\partial M}{\partial S}\right)_P=\frac{3\Lambda r_H^5-2\alpha\Lambda r_H^4-3r_H^3+3(\alpha+\beta\Lambda)r_H^2-6\beta}{6\pi\alpha r_H^3-12\pi r_H^4-6\pi\beta r_H}.
\end{equation}
\begin{equation}
	V=\mathcal{W}\left(\frac{\partial M}{\partial P}\right)_S=\frac{4\pi r_H^3}{3}.
\end{equation}
This is consistent with the results we obtained using general black hole thermodynamics.

Next, let's study criticality by introducing the concept of specific volume $v=2r_H$ \cite{kubizvnak2012p, wei2015insight}. By using (\ref{eq:4}) and (\ref{eq:5}), we can obtain the equation of state:
\begin{equation}
	P=\frac{6\alpha v^2-3v^3+48\beta+6\pi Tv^4+24\pi\beta Tv-6\pi\alpha Tv^3}{6\pi v^5-8\pi\alpha v^4+48\pi\beta v^2}.
\end{equation}
The critical point must meet:
\begin{equation}
	\frac{\partial^2P}{\partial v^2}=\frac{\partial P}{\partial v}=0.
\end{equation}
For three different high-order correction cases of $\beta=\alpha/10$, $\beta=\alpha$, and $\beta=10\alpha$, the critical ratio is:
\begin{equation}
	\left.\frac{P_cv_c}{T_c}\right|_{\beta=0}=0.3667.
\end{equation}
\begin{equation}
	\left.\frac{P_cv_c}{T_c}\right|_{\beta=\frac{\alpha}{10}}=0.3732.
\end{equation}
\begin{equation}
	\left.\frac{P_cv_c}{T_c}\right|_{\beta=\alpha}=0.3856.
\end{equation}
\begin{equation}
	\left.\frac{P_cv_c}{T_c}\right|_{\beta=10\alpha}=0.3937.
\end{equation}
It can be clearly seen that as higher-order corrections dominate, the critical ratio will increase, surpassing Van der Waals system (0.375).

Finally, in order to better illustrate that the thermodynamic behavior of high-order modified Schwarzschild-AdS black hole will return to uncorrected case, we calculated the black hole's thermal capacity and Gibbs free energy. Since $S\propto V^{2/3}$, the isochoric heat capacity must be zero, that is:
\begin{equation}
	C_V=T\left(\frac{\partial S}{\partial T}\right)_V=0.
\end{equation}
On the other hand, the isobaric heat capacity is:
\begin{equation}
	C_P=T\left(\frac{\partial S}{\partial T}\right)_P=\frac{2\pi r_H^2F(r_H^3G-3\alpha r_H^2+6\beta H)}{6r_H^6A+12\alpha r_H^5B+\alpha^2r_H^4C-42\beta r_H^3+3\alpha\beta r_H^2D+6\beta^2E}.
\end{equation}
Where $A=8\pi Pr_H^2-1$, $B=1-4\pi Pr_H^2$, $C=16\pi r_H^2-3$, $D=5-8\pi Pr_H^2$, $E=4\pi Pr_H^2-1$, $F=\beta+r_H^2(2r_H-\alpha)$, $G=3+8\pi Pr_H(3r_H-2\alpha)$, $H=1+4\pi Pr_H^2$.

\begin{figure}
	\centering
	\subfloat[The heat capacity $C_P$ at $\alpha=0$, $\beta=0$.]
	{\includegraphics[width=6cm]{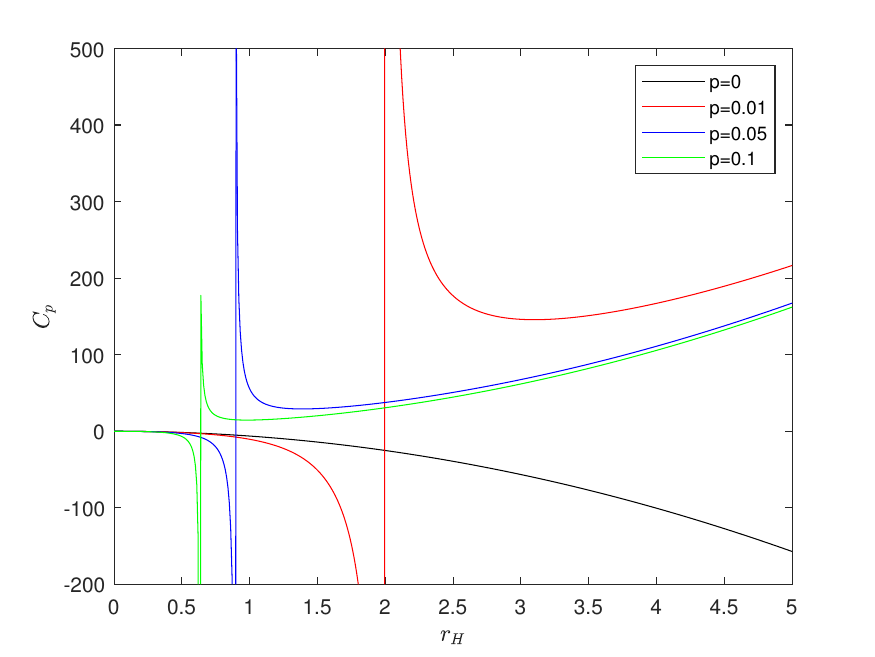}\label{fig:subfig1}}\quad
	\subfloat[The heat capacity $C_P$ at $\alpha=0.1$, $\beta=0$.]
	{\includegraphics[width=6cm]{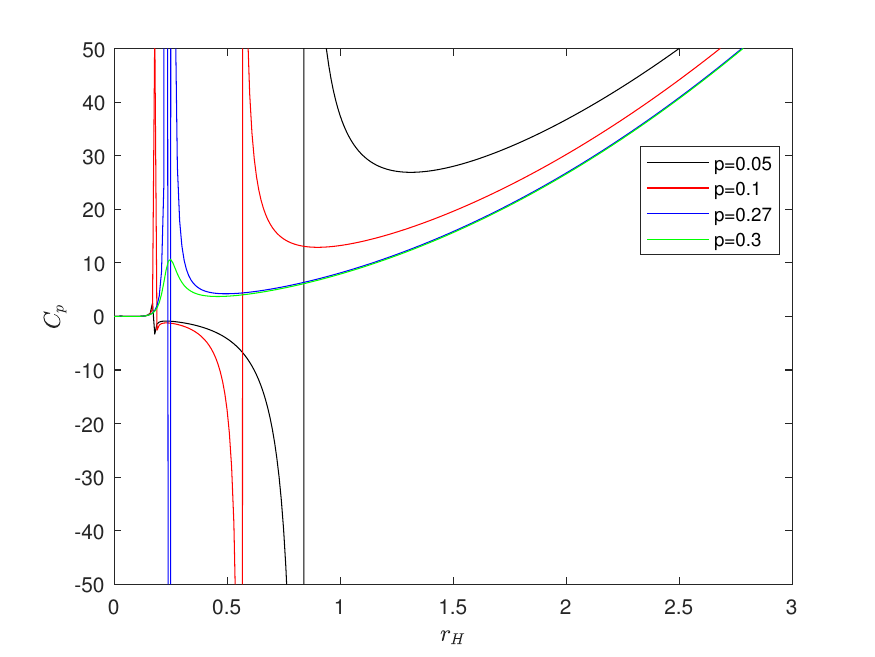}\label{fig:subfig2}}\quad
	\subfloat[The heat capacity $C_P$ at $\alpha=0.1$, $\beta=0.01$.]
	{\includegraphics[width=6cm]{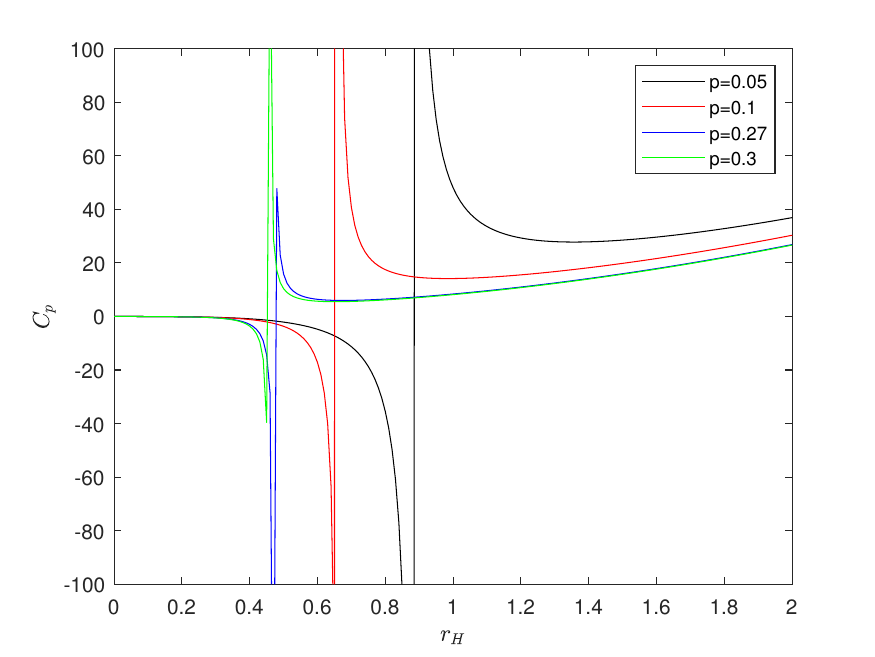}\label{fig:subfig3}}\quad
	\subfloat[The heat capacity $C_P$ at $\alpha=0.1$, $\beta=0.1$.]
	{\includegraphics[width=6cm]{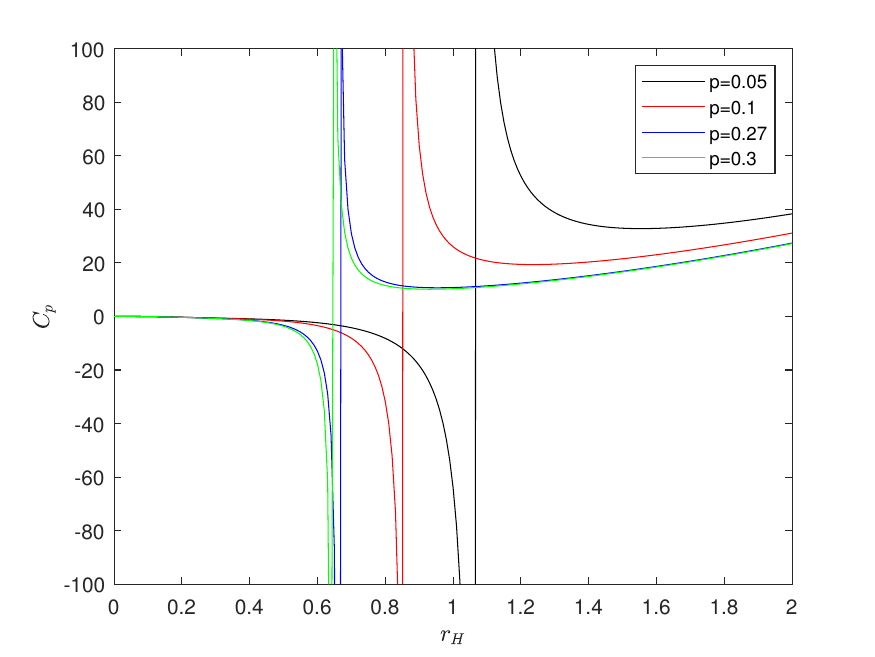}\label{fig:subfig4}}\quad
	\subfloat[The heat capacity $C_P$ at $\alpha=0.1$, $\beta=1$.]
	{\includegraphics[width=10cm]{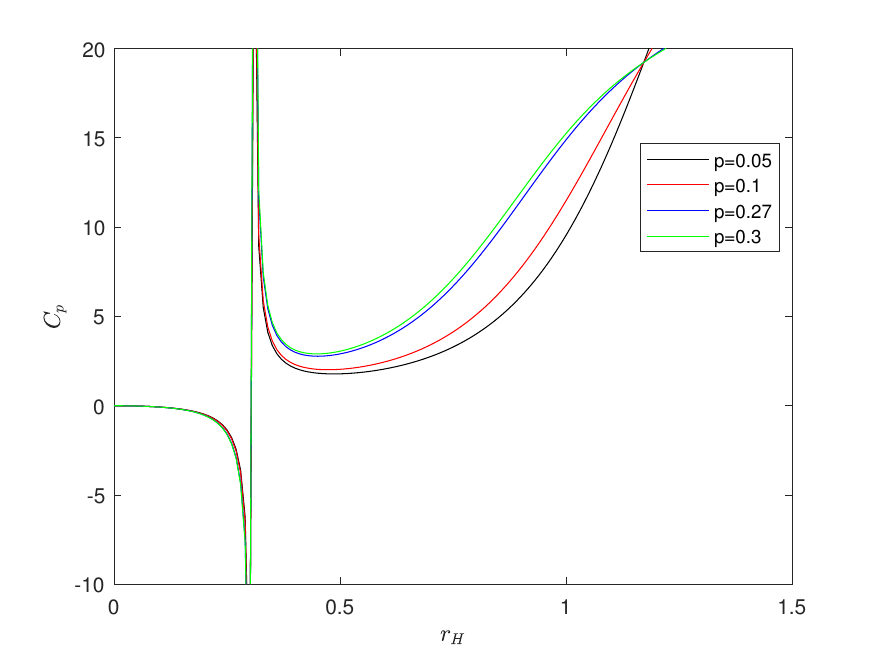}\label{fig:subfig5}}
	\caption{\label{fig:3}The heat capacity $C_P$ as a function of $r_H$ for different pressure.}
\end{figure}
We have plotted the isobaric heat capacities under three different scenarios: without correction, with leading-order correction, and with higher-order correction, in figure \ref{fig:3}. Among them, figure \ref{fig:subfig1} shows the case without correction ($\alpha=0$, $\beta=0$), \ref{fig:subfig2} shows the case with only leading order correction($\alpha=0.1$, $\beta=0$), and figure \ref{fig:subfig3} \ref{fig:subfig4} \ref{fig:subfig5} shows the case with high-order correction, figure \ref{fig:subfig3}: $\alpha=0.1$, $\beta=0.01$, figure \ref{fig:subfig4}: $\alpha=0.1$, $\beta=0.1$, figure \ref{fig:subfig5}: $\alpha=0.1$, $\beta=1$, respectively. It can be clearly seen from figure \ref{fig:3} that for the case where only leading order correction is considered, there are two phase transition points when $P<P_c$, and the phase transition point disappears when $P>P_c$. However, for the case without correction and with high-order correction, regardless of how $P$ changes, there is only one phase transition point.

According to the definition of Gibbs free energy, we have:
\begin{equation}
	G=M-TS=\frac{8\pi Pr_H^6-16\pi\alpha Pr_H^5-3r_H^4-3(\alpha-8\pi\beta P)r_H^3+6\beta r_H}{6\alpha r_H^2-12r_H^3-6\beta}.\label{eq:6}
\end{equation}
According to classical thermodynamics, Gibbs free energy should satisfy:
\begin{equation}
	\mathrm{d}G=-S\mathrm{d}T+V\mathrm{d}P.
\end{equation}
For non-commutative geometry, considering the correction of first law of thermodynamics, the differential of Gibbs free energy should also be corrected to:
\begin{equation}
	\mathrm{d}G=(\mathcal{W}^{-1}-1)T\mathrm{d}S+\mathcal{W}^{-1}V\mathrm{d}P-S\mathrm{d}T.\label{eq:7}
\end{equation}
In the temperature monotonic region of Gibbs free energy, (\ref{eq:7}) can be rewritten as:
\begin{equation}
	\mathrm{d}G=[-S+(\mathcal{W}^{-1}-1)C_P]\mathrm{d}T+\left[\mathcal{W}^{-1}V-(\mathcal{W}^{-1}-1)C_P\left(\frac{\partial T}{\partial P}\right)_S\right]\mathrm{d}P.
\end{equation}
We can further obtain:
\begin{equation}
	\frac{\partial G}{\partial T}=-S+(\mathcal{W}^{-1}-1)C_P.
\end{equation}
\begin{equation}
	\frac{\partial G}{\partial P}=\mathcal{W}^{-1}V-(\mathcal{W}^{-1}-1)C_P\left(\frac{\partial T}{\partial P}\right)_S.
\end{equation}

\begin{figure}
	\centering
	\subfloat[The Gibbs free energy $G$ at $\alpha=0$, $\beta=0$.]
	{\includegraphics[width=6cm]{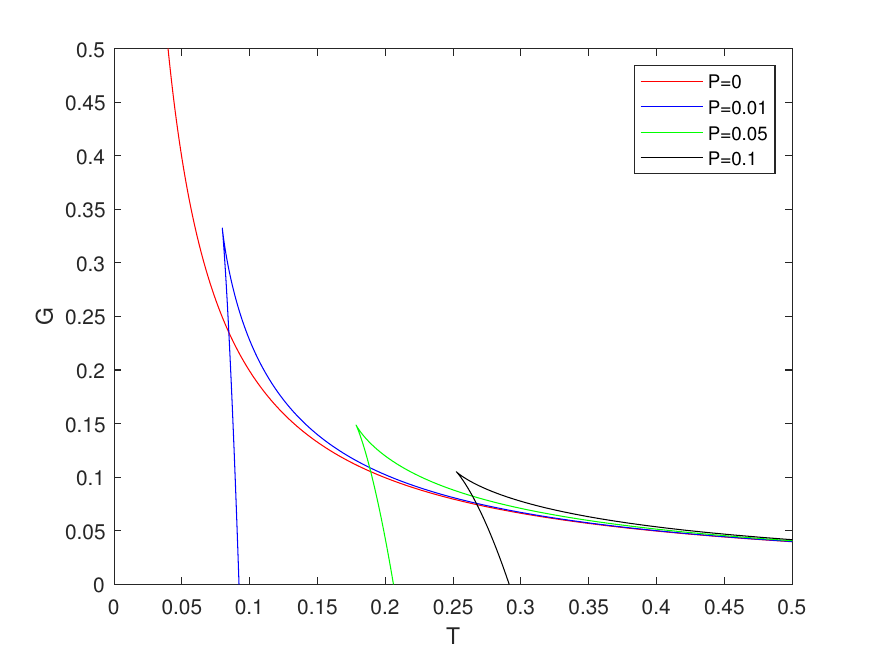}\label{fig:subfig6}}\quad
	\subfloat[The Gibbs free energy $G$ at $\alpha=0.1$, $\beta=0$.]
	{\includegraphics[width=6cm]{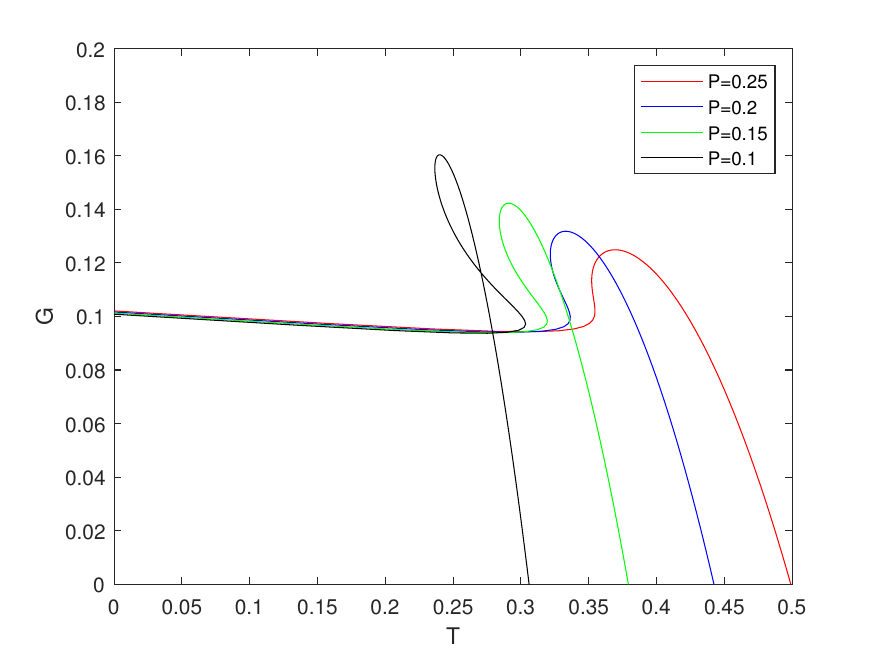}\label{fig:subfig7}}\quad
	\subfloat[The Gibbs free energy $G$ at $\alpha=0.1$, $\beta=0.01$.]
	{\includegraphics[width=6cm]{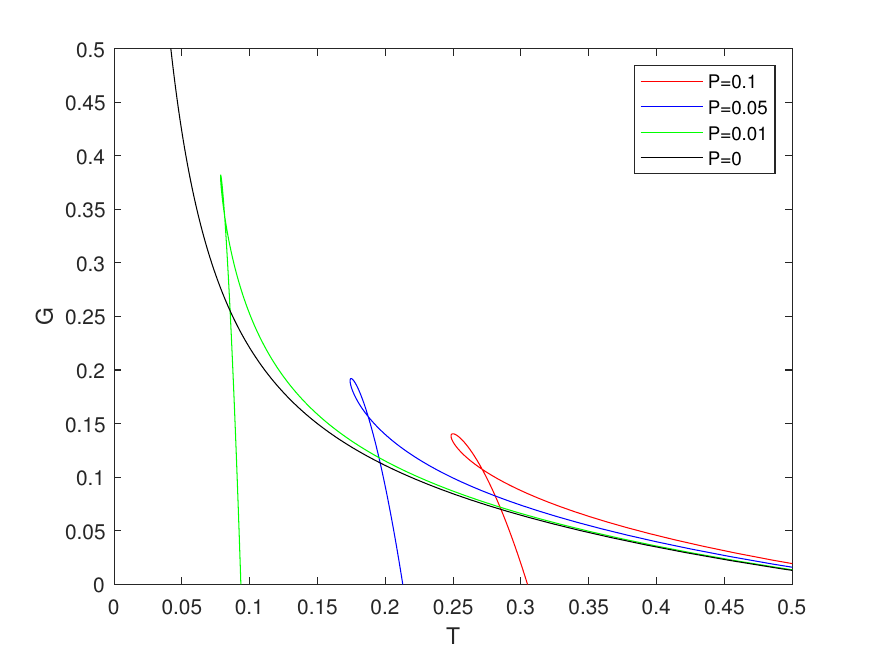}\label{fig:subfig8}}\quad
	\subfloat[The Gibbs free energy $G$ at $\alpha=0.1$, $\beta=0.1$.]
	{\includegraphics[width=6cm]{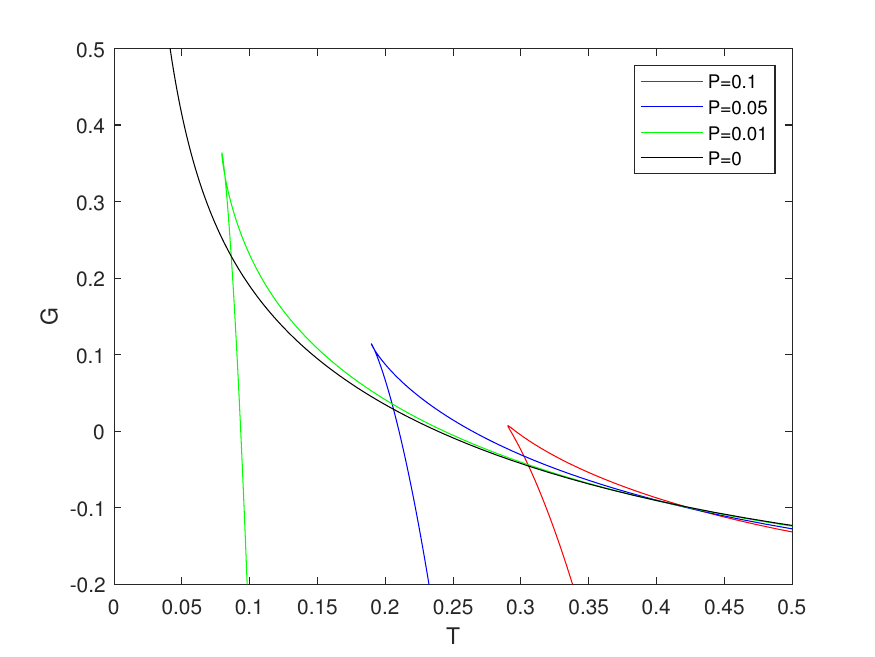}\label{fig:subfig9}}\quad
	\subfloat[The Gibbs free energy $G$ at $\alpha=0.1$, $\beta=1$.]
	{\includegraphics[width=6cm]{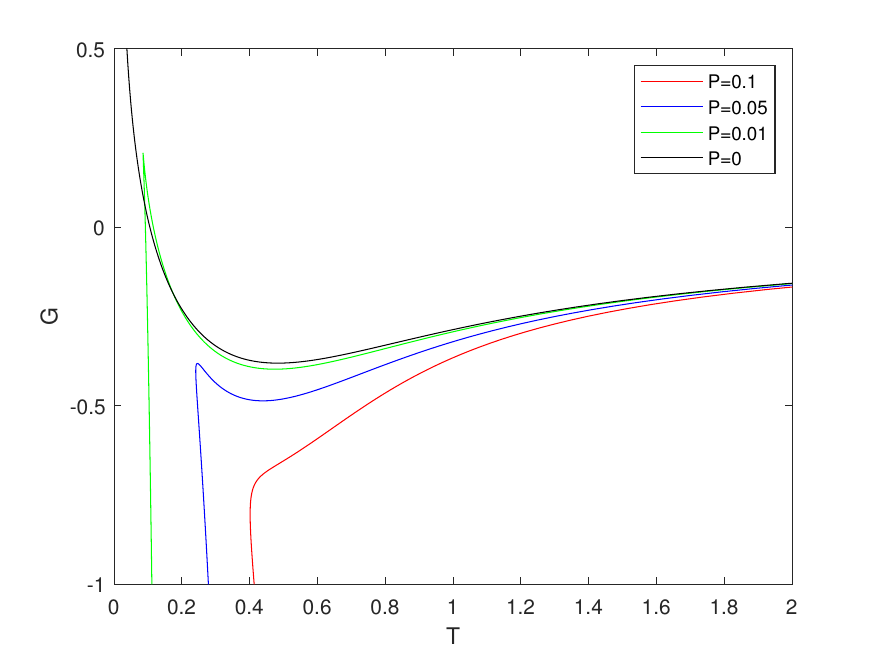}\label{fig:subfig10}}\quad
	\subfloat[The Gibbs free energy $G$ at $\alpha=0.1$, $\beta=10$.]
	{\includegraphics[width=6cm]{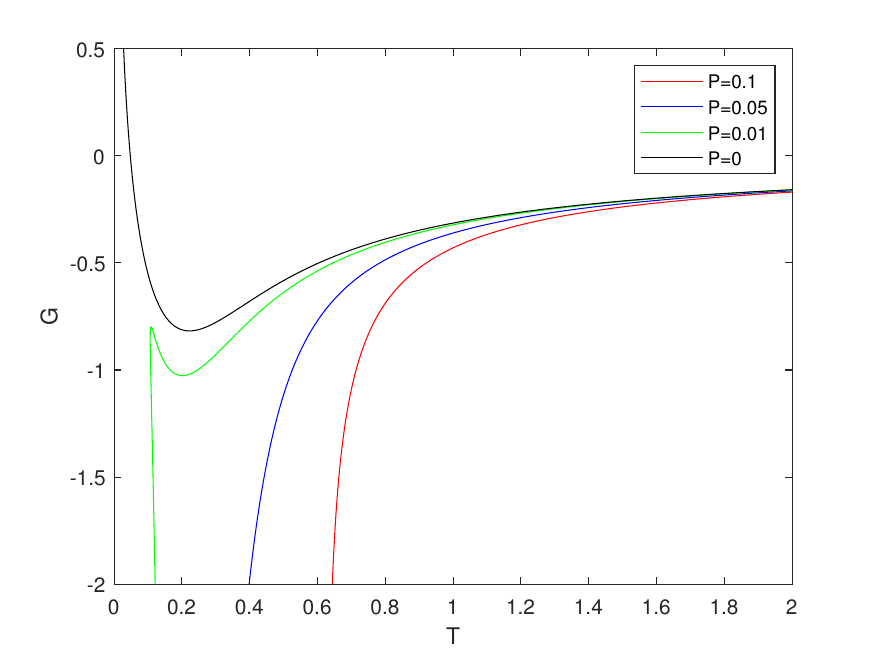}\label{fig:subfig11}}
	\caption{\label{fig:4}The Gibbs free energy $G$ as a function of temperature $T$ for different pressure.}
\end{figure}
We can obtain the relationship between Gibbs free energy and temperature based on (\ref{eq:5}) and (\ref{eq:6}) and draw it in figure \ref{fig:4}. Among them, figure \ref{fig:subfig6} shows the case without correction ($\alpha=0$, $\beta=0$), figure \ref{fig:subfig7} shows the case with only leading order correction ($\alpha=0.1$, $\beta=0$), and figure \ref{fig:subfig8} \ref{fig:subfig9} \ref{fig:subfig10} \ref{fig:subfig11} shows the case with high-order correction, figure \ref{fig:subfig8}: $\alpha=0.1$, $\beta=0.01$, figure \ref{fig:subfig9}: $\alpha=0.1$, $\beta=0.1$, figure \ref{fig:subfig10}: $\alpha=0.1$, $\beta=1$, figure \ref{fig:subfig11}: $\alpha=0.1$, $\beta=10$, respectively. It can be clearly seen from figure \ref{fig:4} that Gibbs free energy of Figure \ref{fig:subfig7} intersects with itself significantly, while Gibbs free energy of Figure \ref{fig:subfig8} intersects slightly with itself. The Gibbs free energies of other figures do not intersect with themselves. Furthermore, we note that when $\alpha=0$, $\beta=0$, the critical ratio is 0.5, when $\alpha=0.1$, $\beta=0$, the critical ratio is 0.3667, when $\alpha=0.1$, $\beta=0.01$, the critical ratio is 0.3732, when $\alpha=0.1$, $\beta=0.1$, the critical ratio is 0.3856, when $\alpha=0.1$, $\beta=1$, the critical ratio is 0.3937, while the critical ratio of Van der Waals system is 0.375. Based on this, we speculate that thermodynamic systems with critical ratios greater than 0.375 and those with critical ratios less than 0.375 have completely different thermodynamic behaviors. Therefore, the thermodynamic behavior of Schwarzschild-AdS black hole under the non-commutative geometric framework considering higher-order corrections will be very similar to the case without corrections, but significantly different from the case considering only leading-order corrections.
\section{Joule-Thomson effect}
\label{sec:3}
In the thermodynamics of black hole, especially in AdS spacetime, researchers have conducted extensive research on the Joule-Thomson effect \cite{bi2021joule, li2020joule, guo2020joule, mo2020effects, okcu2018joule, mo2018joule, lan2018joule, okcu2017joule}. The Joule-Thomson process is an isentropic process. The mass of an AdS black hole is considered as enthalpy, so studying the isomass process of black holes is of great importance. To study isomass process, we rewrite temperature and pressure of the black hole with higher-order corrections as functions of mass $M$ and event horizon radius $r_H$:
\begin{equation}
	P=\frac{6Mr_H^3-3\alpha Mr_H^2+3\beta M-3r_H^4}{8\pi r_H^6}.
\end{equation}
\begin{equation}
	T=\frac{3Mr_H^3-2\alpha Mr_H^2+3\beta M-r_H^4}{2\pi r_H^5}.
\end{equation}
So we obtained the high-order corrected Joule-Thomson coefficient:
\begin{equation}
	\mu=\left(\frac{\partial T}{\partial P}\right)_M=\frac{1}{C_p}\left[T\left(\frac{\partial V}{\partial T}\right)_P-V\right]=\frac{2r_H^5-12Mr_H^4+12\alpha Mr_H^3-30\beta Mr_H}{3r_H^4-9Mr_H^3+6\alpha Mr_H^2-9\beta M}.
\end{equation}

\begin{figure}
	\centering
	\includegraphics[width=10cm]{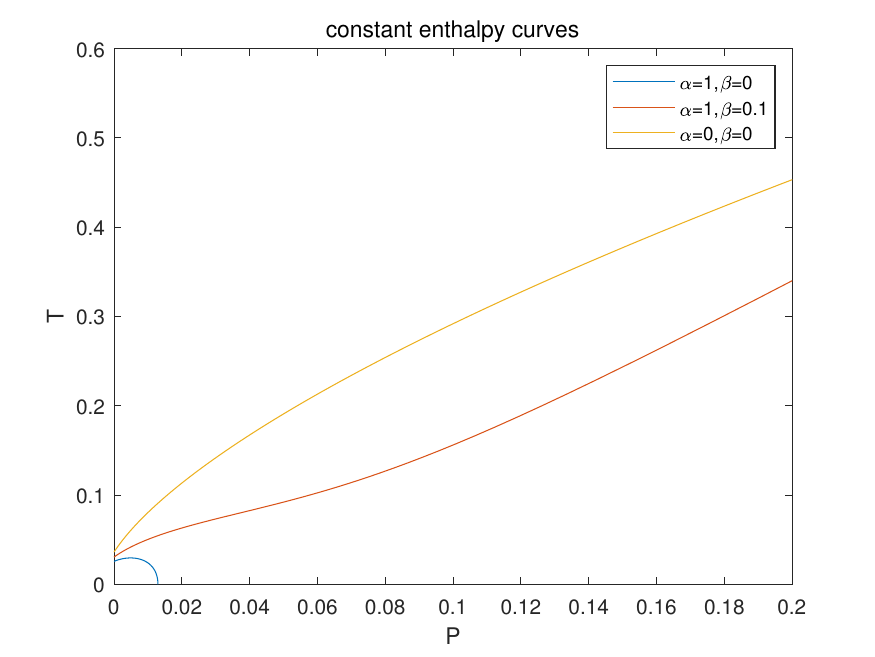}
	\caption{\label{fig:1}The constant mass curves at $\alpha=0$, $\beta=0$ (yellow line), $\alpha=1$, $\beta=0$ (blue line), $\alpha=1$, $\beta=0.1$ (red line). We set $M=1.1$.}
\end{figure}
The process of $\mu>0$ is called cooling process, while the process of $\mu<0$ is called heating process. The point at $\mu=0$ is called inversion point. The curve formed by inversion point in the P-T graph is inversion curve. $\mu=0$ can defines the inversion temperature:
\begin{equation}
	T_i=V\left(\frac{\partial T}{\partial V}\right)_P=\frac{(8\pi\beta^2r_H^2-40\pi\alpha\beta r_H^4+96\pi\beta r_H^5+16\pi\alpha^2r_H^6-112\pi\alpha r_H^7+144\pi r_H^8)P+6r_H^6-\alpha^2r_H^4-18\beta r_H^3+17\alpha\beta r_H^2-6\beta^2}{18\pi r_H(\alpha r_H^2-2r_H^3-\beta)^2}.\label{eq:8}
\end{equation}
From (\ref{eq:5}), we have:
\begin{equation}
	T_i=\frac{(16\pi\alpha r_H^4-24\pi r_H^5-24\pi\beta r_H^2)P+3\alpha r_H^2-3r_H^3-6\beta}{6\pi r_H(\alpha r_H^2-2r_H^3-\beta)}.\label{eq:9}
\end{equation}
According to (\ref{eq:8}) and (\ref{eq:9}), we can get a positive and real root $r_H(P_i)$. Where $P_i$ is the inversion pressure. Then substituting the root into (\ref{eq:8}), we can obtain the inversion temperature $T_i$. Since these expressions are too complex, it is not specifically written here. The inversion point is also extreme value point of tangent slope of the isenthalpic line. We take $M=1.1$ as an example to draw the isenthalpic lines in figure \ref{fig:1} for three cases of no correction ($\alpha=0$, $\beta=0$), only leading order correction ($\alpha=1$, $\beta=0$), and high-order correction ($\alpha=1$, $\beta=0.1$). From figure \ref{fig:1}, it can be seen that the isenthalpic line with high-order correction is very similar to the isenthalpic line without correction, while the case with only leading order correction is very different. This once again confirms our hypothesis that considering high-order corrections, thermodynamic properties of non-commutative Schwarzschild-AdS black hole will return to that of uncorrected case.
\section{Conclusion and discussion}
\label{sec:4}
In this paper, we calculated high-order corrections to thermodynamics of Schwarzschild-AdS black hole within the framework of non-commutative geometry. By solving Einstein's field equation with cosmological constant, we obtained a high-order modified Schwarzschild solution and studied its thermodynamics. Our calculations have verified that the stress-energy tensor containing black hole mass will correct first law of black hole thermodynamics. In addition, our results indicate that the thermodynamic behavior of a high-order corrected Schwarzschild-AdS black hole will be more similar to that of a regular Schwarzschild-AdS black hole, rather than only considering the leading order corrected Schwarzschild-AdS black hole. In addition, we calculated the critical behavior of Schwarzschild-AdS black hole under high-order correction and found that its critical ratio increases when high-order correction plays a major role. Finally, we calculated the Joule-Thomson effect of Schwarzschild-AdS black hole under high-order correction and provided the inversion temperature curve, which enhances our understanding of the thermodynamic behavior of black holes under non-commutative geometry. 

For phenomenological descriptions of quantum black holes, such as the generalized uncertainty principle, non commutative geometry, and modified gravity models, researchers rarely consider higher-order modifications of these effects. The thermodynamics of the Schwarzschild AdS black hole within the framework of noncommutative geometry mentioned in this paper, after taking into account the higher-order corrections, has properties that are more similar to those of an ordinary Schwarzschild AdS black hole. This can serve as a remarkable signature distinguishing noncommutative geometry from other phenomenological models of quantum gravity. It is worth noting that for Einstein-Gauss-Bonnet gravity, when only considering its leading-order correction, it reduces to an ordinary Schwarzschild black hole, or in the case of a charged black hole, it reduces to a Reissner-Nordstr\"{o}m (RN) black hole. The second-order correction will lead to an additional correction term $\frac{a}{r_H^2}$. Where $a$ is Gauss-Bonnet coupling coefficient. Details of Einstein-Gauss-Bonnet gravity can be found in \cite{wu2021hawking, fernandes2020charged, ghosh2020thermodynamic, lekbich20234d, assrary2022effect, hegde2024thermodynamics}. However, the higher-order corrections have not been studied by anyone yet. In future work, we aim to use the AdS-CFT duality to provide a microscopic explanation for the above-mentioned phenomena.

\section*{Acknowledgements}
The authors are grateful to anonymous reviewers for their very crucial and detailed comments. The authors would like to thank Prof. Jian Jing and his student Liubiao Ma and Zheng Wang from the Department of Physics, Beijing University of Chemical Technology for their valuable comments and suggestions during the completion of this manuscript. The author would like to thank Jian-Bo Deng from the Department of Physics, Lanzhou University, whose article \cite{wang2024thermodynamics} provided great inspiration during the process of completing the manuscript.
  \bibliographystyle{elsarticle-num} 
  \bibliography{noncommutative_BH}

\end{document}